\documentclass[12pt,preprint]{aastex}

\shorttitle{Separating the Lensing and KSZ Effects}
\shortauthors{Riquelme \& Spergel}

\begin{document}

\title{Separating the Weak Lensing and Kinetic SZ Effects \\ from CMB Temperature Maps}

\author{Mario A. Riquelme\altaffilmark{1} and David N. Spergel}

\affil{Department of Astrophysical Sciences, Princeton University, Princeton, NJ 08544}

\email{marh@astro.princeton.edu, dns@astro.princeton.edu}

\altaffiltext{1}{Departamento de Astronomia y Astrofisica, P. Universidad Catolica de Chile, where part of this work was realized.}

\begin{abstract}
A new generation of CMB experiments will soon make sensitive high resolution maps of the microwave sky. At angular scales less than $\sim$10 arcminutes, most CMB anisotropies are generated at z $< 1000$, rather than at the surface of last scattering. Therefore, these maps potentially contain an enormous amount of information about the evolution of structure. Whereas spectral information can distinguish the thermal Sunyaev-Zeldovich (tSZ) effect from other anisotropies, the spectral form of anisotropies generated by the gravitational lensing and the kinetic Sunyaev-Zeldovich (kSZ) effects are identical. While spectrally identical, the statistical properties of these effects are different. We introduce a new real-space statistic, $<\theta (\hat{n})^3 \theta (\hat{m})>_c$, and show that it is identically zero for weakly lensed primary anisotropies and, therefore, allows a direct measurement of the kSZ effect. Measuring this statistic can offer a new tool for studing the reionization epoch. Models with the same optical depth, but different reionization histories, can differ by more than a factor of 3 in the amplitude of the kSZ-generated non-Gaussian signal.
\end{abstract}

\keywords{cosmic microwave background --- gravitational lensing --- structure formation}

\section{Introduction}

An upcoming generation of survey telescopes, such as ACT\footnote{http://www.hep.upenn.edu/act/act.html}, APEX\footnote{http://bolo.berkeley.edu/apexsz/}, and SPT\footnote{http://astro.uchicago.edu/spt/}, will map the cosmic microwave background (CMB) with an unprecedented sensitivity and with resolution of $\sim$1' .\newline
\indent On degree scales, the main features of the CMB are fluctuations generated at the last scattering surface (z $\sim$ 1100). However, on smaller angular scales, secondary effects produced through the interaction between the CMB and the large-scale matter distribution of the universe dominate the microwave sky. Consequently, high resolution experiments will open a new window into different physical processes that affect the CMB photons as they travel from the recombination epoch to the present. In particular, these experiments are expected to provide invaluable information about the thermal Sunyaev-Zeldovich (tSZ), the kinetic Sunyaev-Zeldovich (kSZ) effects, and the weak gravitational lensing effects. \newline
\indent The weak lensing effect appears to be a particularly promising probe of the growth of structure. Its imprints on the CMB sky have the potential of providing us with important information about the matter distribution over most of the cosmological history. So, besides testing our paradigm of structure formation, the measurement of the lensing effect can be very useful for constraining parameters like neutrino mass \citep{Kap03} and the dark energy equation-of-state \citep{GolSpe99, Kap03}. Lensing measurements can also assist CMB polarization measurements in detecting primordial gravity waves \citep{SelHir04}.\newline
\indent There has been significant study of the effects of CMB lensing on the temperature and polarization power spectrum. Power spectrum analysis of these quantities can give us important information about the power spectrum of dark matter \citep{Sel96, ZalSel98}. In addition, several works have been devoted to the development of methods for extracting the projected matter field using the higher order correlations induced by lensing on the CMB temperature and polarization \citep{Hu01a,HirSel03a,HuOka02,HirSel03b,OkaHu03}.\newline
\indent Efforts to measure gravitational lensing may be limited by contamination produced by the tSZ and kSZ effects, and also by extragalactic point sources. Whereas these contaminants are thought to be unimportant for polarization maps \citep{Hu01c}, they will likely contaminate the temperature anisotropies at angular scales of $\sim10$ arcminutes and below. The spectral dependence of the tSZ effect and the extragalactic point sources in principle provides the key for their removal from CMB temperature maps. However, the kSZ effect is spectrally indistinguishable from the lensed primary CMB anisotropies, so a different method is required for removing this effect from CMB maps.\newline
\indent A first method for reducing the kSZ contamination of the lensing signal has already been proposed by \cite{AmbValWhi04}. It uses the spatial correlation between the tSZ and kSZ effects to mask the main sources of kSZ anisotropies. Alternatively, in this work we suggest that the different statistical properties of the kinetic SZ and weak lensing effects, in particular the non-Gaussian features that they produce on the CMB anisotropies, can provide another tool for their separation. We introduce a new real-space statistic, $<\theta (\hat{n})^3 \theta (\hat{m})>_c$, that we show vanishes for weakly lensed primary CMB anisotropies. Consequently, this statistic would give us a clean measurement of the kSZ effect. This measurement is expected to be useful for quantifying and controlling the kSZ contamination of lensing field estimators, in addition to provide a new insight into the history of reionization and structure formation of the universe.\newline
\indent This work is organized as follows: In chapter \ref{Lensing} we begin by reviewing the lensing effect on the CMB temperature, and introduce the $<\theta (\hat{n})^3 \theta (\hat{m})>_c$  statistic, showing how it vanishes for weakly lensed primary anisotropies. In chapter \ref{kSZ} we review the kSZ effect and illustrate, using the Ostriker-Vishniac (OV) effect, how $<\theta^3(\hat{n})\theta(\hat{m})>_c$ can be used to determine the different physical parameters involved in our models for the kSZ effect. In chapter \ref{Conclusiones} we present our conclusions.

\section{Weak Lensing of the CMB}\label{Lensing}

Weak lensing on the CMB temperature can be described as a mapping of the unlensed
temperature anisotropies, $\theta(\hat{n})$, into the lensed ones,  $\theta_L(\hat{n})$. (Note that we use $\theta(\hat{n}) \equiv T(\hat{n})/\bar{T}_{CMB} - 1$ for describing temperature anisotropies). The relation between these two maps is
\begin{eqnarray}
\theta_L(\hat{n}) = \theta(\hat{n} + \delta \hat{n}),
\label{eq:relation}
\end{eqnarray}
where the unitary vector $\hat{n}$ defines a direction on the sky and $\delta \hat{n}$ represents the lensing deflection angle due to the intervening mass distribution.\newline

\subsection{Weak lensing four-point function}
\label{sec:LenTris}

According to the standard inflationary scenario, the primordial density perturbations on the last scattering surface can be regarded as a nearly Gaussian random field. Consequently, the primary (unlensed) temperature anisotropies of the CMB will also have a Gaussian m-point probability distribution given by,
\begin{eqnarray}
P(\theta(\hat{n}_1),\dots,\theta(\hat{n}_m)) d\theta(\hat{n}_1)\dots d\theta(\hat{n}_m) & = & \frac{1}{\sqrt{(2\pi)^{m}\det(M)}}\exp\big[ -\frac{1}{2} \sum_{i,j=1}^m \theta(\hat{n}_i) (M^{-1})_{ij} \theta(\hat{n}_j) \big] \nonumber \\ && \times d\theta(\hat{n}_1)\dots d\theta(\hat{n}_m),
\label{eq:gaussian}
\end{eqnarray}
where $m$ is an arbitrary positive integer and $M$, the covariance matrix, is such that $M_{ij} = <\theta(\hat{n}_i)\theta(\hat{n}_j)>$. Therefore, the statistical nature of the unlensed anisotropies is completely specified by the two-point correlation function. More precisely, according to the Wick's theorem, any $m$-point correlation function for a Gaussian field will be given by,
\begin{eqnarray}
<\theta(\hat{n}_1)\dots \theta(\hat{n}_{m})> = \left\{ \begin{array}{ll}
\sum_{\textrm{pairs}} <\theta(\hat{n}_1)\theta(\hat{n}_{2})>\dots <\theta(\hat{n}_{m-1})\theta(\hat{n}_{m})> & \textrm{ for m even } \\
0 & \textrm{ for m odd, } \end{array} \right.
\end{eqnarray}
where $\sum_{\textrm{pairs}}$ means that the sum must be performed over the $(m-1)\cdot(m-3)\dots3\cdot1$ ways to make pairs.\newline
\indent Therefore, the appearance of an additional term, usually called connected part, in the $m$-point function is a direct probe of non-Gaussianities that could have been induced by secondary effects like lensing or the kSZ effects. Since both effects are expected to have a null three-point function, in this work we will explore the potential of the connected part of the four-point function for accomplishing the separation of these two effects.\newline
\indent In general, weak lensing produces a non-zero connected part for the four-point function of CMB temperature. It can be seen by Taylor expanding equation (\ref{eq:relation}),
\begin{eqnarray}
\theta_L(\hat{n}) \approx \theta(\hat{n})  + \partial_i \theta(\hat{n})(\delta \hat{n})_i + \frac{1}{2} \partial_i \partial_j \theta(\hat{n}) (\delta \hat{n})_i (\delta \hat{n})_j,
\label{eq:expansion}
\end{eqnarray}
and calculating the first correction to the connected part of the four-point function due to lensing. Doing so one obtains,
\begin{eqnarray}
\begin{array}{lll}
<\theta_L(\hat{n}_1) \theta_L(\hat{n}_2) \theta_L(\hat{n}_3)\theta_L(\hat{n}_4)>_c & = & \frac{1}{2} \sum_{l,m,p,q = 1}^4 <(\delta \hat{n}_l)_i (\delta \hat{n}_m)_j> \times \\
& & <\theta(\hat{n}_p) \partial_i\theta(\hat{n}_l)> <\theta(\hat{n}_q) \partial_j\theta(\hat{n}_m)> \Delta_{lmpq},
\end{array}
\label{eq:lentri}
\end{eqnarray}
where the index ``c'' means ``connected part'', and $\Delta_{lmpq}$ is $1$ if $l$, $m$, $p$, $q$ are all different from each other and $0$ in all other cases.

\subsection{The $ <\theta (\hat{n})^3 \theta (\hat{m})>_c $ statistic}
\label{sec:statistica}

In this section, we introduce a new statistic, $<\theta (\hat{n})^3 \theta (\hat{m})>_c$, and show that it vanishes for the  weakly lensed primary anisotropies. It is defined as
\begin{eqnarray}
<\theta (\hat{n})^3 \theta (\hat{m})>_c & \equiv & <\theta (\hat{n})^3 \theta (\hat{m})> - 3<\theta (\hat{n})^2><\theta (\hat{n}) \theta (\hat{m})>.
\label{eq:statistic}
\end{eqnarray}
Notice that, since it is a real-space statistic, it is not affected by any type of masking, and is straightforward to compute from temperature maps.\newline
\indent As an illustration, the cancellation of this statistic for the weak lensing case can be easily seen from equation (\ref{eq:lentri}). If we set $\hat{n}_1 = \hat{n}_2 = \hat{n}_3 = \hat{n}$ and $\hat{n}_4 = \hat{m}$, we obtain,
\begin{eqnarray}
<\theta_L (\hat{n})^3 \theta_L (\hat{m})>_c = 6 (<(\delta \hat{n})_i (\delta \hat{n})_j> + <(\delta \hat{n})_i (\delta \hat{m})_j>) <\theta(\hat{m}) \partial_i\theta(\hat{n})><\theta(\hat{n}) \partial_j\theta(\hat{n})>.
\label{eq:caca}
\end{eqnarray}
But the last factor of equation (\ref{eq:caca}) is $<\theta(\hat{n}) \partial_j\theta(\hat{n})> = \frac{1}{2}<\partial_j \theta(\hat{n})^2> = \frac{1}{2}\partial_j <\theta(\hat{n})^2> = 0$, provided that $<\theta(\hat{n})^2>$ does not depend on $\hat{n}$. This way we show that, to the lowest order in $ \delta \hat{n} $, $<\theta_L (\hat{n})^3 \theta_L (\hat{m})>_c = 0$. Eventhough equation (\ref{eq:lentri}) is a good approximation (since most of the lensing effect on the CMB is produced by linear density fluctuations, in which case $\vert \delta \hat{n} \vert$ is much smaller than the angular size of primary CMB anisotropies \citep{Sel96}), we provide below a more general demonstration.\newline
We know that the CMB anisotropies can be expanded into spherical harmonics,
\begin{eqnarray} 
\theta_L(\hat{n}) = \theta(\hat{n} + \delta \hat{n}) = \sum_{lm} a_{lm} Y_l^m(\hat{n} + \delta \hat{n}),
\label{eq:spherharmonic}
\end{eqnarray}
where $\sum_{lm}$ represents $\sum_{l=0}^{\infty}\sum_{m=-l}^l$. Then,
\begin{eqnarray}
<\theta_L (\hat{n})^3 \theta_L (\hat{m})> & = & \sum_{l_1m_1} \sum_{l_2m_2} \sum_{l_3m_3} \sum_{l_4m_4} <Y_{l_1}^{m_1}(\hat{n} + \delta \hat{n})Y_{l_2}^{m_2}(\hat{n} + \delta \hat{n}) Y_{l_3}^{m_3}(\hat{n} + \delta \hat{n})Y_{l_4}^{m_4}(\hat{m} + \delta \hat{m})> \nonumber \\ && \times  <a_{l_1m_1}a_{l_2m_2}a_{l_3m_3}a_{l_4m_4} >.
\label{eq:spherharmonic2}
\end{eqnarray}
\noindent Since we assume that the primary CMB anisotropies are a Gaussian random field, the last term in equation (\ref{eq:spherharmonic2}) can be decomposed into the sum of three terms corresponding to the three ways to make pairs. Thus, defining $<a_{lm}\bar{a}_{l'm'}> \equiv \delta_{ll'}^k \delta_{mm'}^k C_l$ where $\delta_{ll'}^k$ and $\delta_{mm'}^k$ correspond to Kronecker deltas and the bar over $a_{l'm'}$ represents the complex conjugate of $a_{l'm'}$, it is easy to show that,
\begin{eqnarray}
<\theta_L (\hat{n})^3 \theta_L (\hat{m})> & = & 3\sum_{lm} \sum_{l'm'} C_l C_{l'}<Y_{l}^{m}(\hat{n} + \delta \hat{n})\bar{Y}_{l}^{m}(\hat{n} + \delta \hat{n}) Y_{l'}^{m'}(\hat{n} + \delta \hat{n})\bar{Y}_{l'}^{m'}(\hat{m} + \delta \hat{m})>
\nonumber \\ & = & 3\sum_{l=0}^{\infty} \sum_{l'=0}^{\infty} C_l C_{l'} <\sum_{m=-l}^{l}Y_{l}^{m}(\hat{n} + \delta \hat{n})\bar{Y}_{l}^{m}(\hat{n} + \delta \hat{n}) \nonumber \\ && \times \sum_{m'=-l'}^{l'}Y_{l'}^{m'}(\hat{n} + \delta \hat{n})\bar{Y}_{l'}^{m'}(\hat{m} + \delta \hat{m})>.
\label{eq:spherharmonic3}
\end{eqnarray} 
But, making use of the addition theorem for spherical harmonics, 
\begin{eqnarray}
\sum_{m=-l}^{l}Y_{l}^{m}(\hat{n}+\delta \hat{n})\bar{Y}_{l}^{m}(\hat{m}+\delta \hat{m}) = \frac{2l + 1}{4\pi} P_l(\cos(\gamma)),
\label{eq:sumtheorem}
\end{eqnarray} 
where $P_l(x)$ corresponds to the Legendre polynomial and $\gamma$ is the angle between $\hat{n}+\delta \hat{n}$ and $\hat{m}+\delta \hat{m}$, and also using that $P_l(\cos(0)) = 1$, we can see that the sum $\sum_{m=-l}^{l}$ in equation (\ref{eq:spherharmonic3}) will be just a function of $l$ and will not depend on the random field $\delta \hat{n}$, so $<\theta_L (\hat{n})^3 \theta_L (\hat{m})>$ can be separated into the product of two sums,
\begin{eqnarray}
<\theta_L (\hat{n})^3 \theta_L (\hat{m})> = 3\sum_{l=0}^{\infty} C_l \frac{2l+1}{4\pi}\sum_{l'=0}^{\infty} C_{l'} \frac{2l'+1}{4\pi}<P_{l'}(\cos(\gamma))>. 
\label{eq:casidem}
\end{eqnarray}
But, using the definitions of $a_{lm}$, $C_l$, and the addition theorem for spherical harmonics, it is easy to see that,
\begin{eqnarray}
<\theta_L (\hat{n}) \theta_L (\hat{m})> = \sum_{l=0}^{\infty} C_l \frac{2l+1}{4\pi}<P_l(\cos(\gamma))>, 
\label{eq:dem}
\end{eqnarray} 
so equation (\ref{eq:casidem}) can be written,
\begin{eqnarray}
<\theta_L (\hat{n})^3 \theta_L (\hat{m})> = 3<\theta_L (\hat{n})^2><\theta_L (\hat{n}) \theta_L (\hat{m})>.
\label{eq:dem2}
\end{eqnarray}
This way we demonstrate that,
\begin{eqnarray}
<\theta_L (\hat{n})^3 \theta_L (\hat{m})>_c = <\theta_L (\hat{n})^3 \theta_L (\hat{m})> - 3<\theta_L (\hat{n})^2><\theta_L (\hat{n}) \theta_L (\hat{m})> = 0.
\label{eq:dem3}
\end{eqnarray}
\indent It is important to keep in mind that the only approximation used in our calculations was to neglect any possible non-Gaussianity of the primary CMB anisotropies. We made no assumptions about the lensing deflection angle, $ \delta \hat{n} $. Because of its cancellation, this statistic has the potential of giving us a direct measurement of the kSZ effect. 

\section{The kinetic Sunyaev-Zeldovich effect}\label{kSZ}

The scattering of CMB photons by free electrons within galaxies and the intergalactic medium creates anisotropies on the CMB temperature. When these anisotropies are due to the thermal energy that the electrons transfer to the CMB photons, the effect is called thermal Sunyaev-Zeldovich (tSZ) effect. This effect modifies the thermal nature of the spectrum of the CMB, reducing the number of low-energy photons and increasing the number of high-energy photons. At nearly 218 GHz the energy spectrum of the CMB remains almost unchanged, giving rise to the so called thermal Sunyaev-Zeldovich ``null", which is very important for identifying this effect from other secondary CMB signals.\newline\indent
On the other hand, if the free electrons have a bulk peculiar motion, additional temperature fluctuations are generated. This effect, known as the kinetic Sunyaev-Zeldovich (kSZ) effect, can be understood considering that in the rest frame of electrons the CMB is seen as hotter in one direction and colder in the opposite direction (Doppler effect). However, after their scattering with the electrons, part of the CMB photons are reemited isotropically in the rest frame of electrons generating a dipolar anisotropy in the rest frame of the CMB. This temperature anisotropy, which does not depend on frequency, is given by \citep[see, for example,][]{Sca00}
\begin{eqnarray}
\theta (\hat{n}) = -\frac{1}{c} \int_0^{\eta_{rc}}d\eta
a(\eta)\sigma_T n_e(\hat{n}\eta,
\eta)e^{-\tau(\eta)} \hat{n} \cdot \textbf{v}( \hat{n}\eta,
\eta),
\label{eq:kSZ}
\end{eqnarray}
where $\eta$ represents the comoving distance along the path of CMB photons, which we will also use as a time coordinate, the subscript ``rc'' means ``at recombination'', $a(\eta)$ is the scale factor, $\textbf{v}( \textbf{r}, \eta)$ corresponds to the bulk velocity of electrons (note that throught this work bold letters will represent three-dimensional vectors, while two-dimensional vectors will be represented by symbols like $\vec{v}$), $c$ is the speed of light, $n_e( \textbf{r}, \eta)$ is the electron density, $\sigma_T$ is the cross section for Thomson scattering, and $\tau(\eta) = \int_0^{\eta} d\eta' \sigma_T \bar{n}_e(\eta') a(\eta')$ is the optical depth to electron scattering at $\eta$. Notice that in equation (\ref{eq:kSZ}), like throughout this paper, we have assumed a flat universe.\newline
\indent Since in this work we are interested in fluctuations at arcminutes scales, we can describe the kSZ effect by using the flat-sky approximation. Then, we will use $\vec{x}$ instead of $\hat{n}$ to refer to the angular position on the sky, where $\hat{n}$ will be assumed very close to the the $\hat{z}$ axis (i.e., $\hat{n} \approx (x_1, x_2, 1)$ where $x_i \ll 1$).\newline
\indent The kSZ effect is usually divided into two effects: the Ostriker-Vishniac (OV) effect and the non-linear kSZ effect. The OV effect is due to the presence of reionized matter within linearly evolving density fluctuations, while the non-linear kSZ effect is produced by electrons within non-linear structures. Given the highly predictive power of linear theory, the OV effect can give us accurate information about reionization, mainly in its initial phase. However, since in the low-redshift universe most of the reionized matter resides in high density objects, a good understanding of both regimes is essential in order to have a complete picture of the kSZ effect.\newline

\subsection{The Ostriker-Vishniac effect}
\label{sec:OV}

In this work, we will focus only on the linear version of the kSZ effect, the OV effect. We will assume that the spatial distribution of free electrons is the same as the spatial distribution of dark matter, i. e., we will assume a homogeneously reionized universe and that dark matter traces baryons. Therefore, we will be interested in $\delta ( \textbf{r},\eta) \equiv \rho(\hat{n}\eta,\eta)/\bar{\rho}(\eta) - 1$ where $\rho$ represents the dark matter density.\newline
\indent In the linear regime, the time evolution of the dark matter density perturbations is given by
\begin{eqnarray}
\delta( \hat{n}\eta, \eta) = \delta( \hat{n}\eta) \frac{D(\eta)}{D(0)},
\end{eqnarray}
where $\delta( \hat{n}\eta) = \delta( \hat{n}\eta, \eta=0)$,
\begin{eqnarray}
D(\eta) = \frac{5\Omega_m H(\eta)}{2c} \int_{\eta}^{\eta_\infty} d \eta' \frac{H_0^2}{a(\eta')H(\eta')^2}
\end{eqnarray}
is the growth function, and $\eta_\infty = \eta (z =\infty)$.\newline
\indent It is useful to define the power spectrum $P(k)$ of the density perturbations as
\begin{eqnarray}
<\widetilde{\delta}(\textbf{k})\widetilde{\delta}(\textbf{k}')^*> \equiv (2\pi)^3 P(k)\delta_D(\textbf{k} - \textbf{k}'),
\end{eqnarray}
where $\widetilde{\delta}(\textbf{k})$ and $\delta_D(\textbf{k})$ correspond to the Fourier transform of $\delta( \textbf{r})$ and the Dirac delta, respectively.\newline
\indent In linear theory, the velocity field can be obtained from the continuity equation. Its Fourier transform is
\begin{eqnarray}
\widetilde{\textbf{v}}(\textbf{k}, \eta) =
\frac{ia(\eta)\dot{D}(\eta)}{k^2D(0)}\tilde{\delta}(\textbf{k})\textbf{k},
\label{eq:continuity}
\end{eqnarray}
where $\dot{D}(\eta)$ denotes the derivative of $D(\eta)$ with respect to {\it time}.\newline
\indent From equation (\ref{eq:continuity}) we see that only Fourier modes with a non-zero line of sight component (i.e., $k_z \neq 0$ in the flat-sky approximation) contribute to $v_z(\textbf{r}, \eta)$. However, for a slow enough reionization process, the successive troughs and crests of these modes tend to cancel when projected along the $\hat{z}$ axis. Because of this, as we will see below, the OV effect arises from the modulation of the velocity field by the spatial fluctuations in the electron density. These fluctuations may be due to inhomogeneities in the baryon density and/or in the ionization fraction. Although the latter case, known as the patchy reionization effect, can give rise to important temperature anisotropies \citep{San03}, we will concentrate in a homogeneously reionized universe where the OV effect is due only to fluctuations in the baryon density.\newline
\indent Then, in the flat-sky approximation, the temperature fluctuations can be written as
\begin{eqnarray}
\theta_{ov} (\vec{x}) = -\frac{1}{c} \int_0^{\eta_{rc}}d\eta g(\eta) q_z(\vec{x}\eta,
\eta, \eta),
\end{eqnarray}
where $g(\eta) \equiv a(\eta)\sigma_T \bar{n}_e(\eta)e^{-\tau(\eta)}$ is called the visibility function, and $\textbf{q}(\textbf{r}, \eta) = \delta (\textbf{r}, \eta)\textbf{v}(\textbf{r}, \eta)$.\newline
The Fourier transform of $\textbf{q}( \textbf{r}, \eta)$ is
\begin{eqnarray}
\widetilde{\textbf{q}}(\textbf{k}, \eta) =\frac{ia\dot{D}D}{D_0^2} \int
\frac{d\textbf{k}'}{(2\pi)^3}\tilde{\delta}( \textbf{k}')\tilde{\delta}(
\textbf{k} - \textbf{k}') \frac{\textbf{k}'}{ k'^2}.
\label{eq:Fourier_q}
\end{eqnarray}
From equation (\ref{eq:Fourier_q}), we can see that $\widetilde{q}_z(\textbf{k},\eta )\vert_{k_z=0 } \neq 0$. This is very important because, given that the modes perpendicular to $\hat{z}$ do not cancel out when projected along the line of sight, in a slowly varying reionization process they dominate the OV effect.\newline
\indent Finally, it can be shown that the Fourier transform of $\theta (\vec{x})$ is
\begin{eqnarray}
\widetilde{\theta}_{ov} (\vec{l}) = -\int_0^{\eta_{rc}}d\eta
\frac{g(\eta)}{c\eta^2} \int \frac{dk_z}{2 \pi}e^{-i k_z \eta} \widetilde{q}_z(\frac{\vec{l}}{\eta}, k_z, \eta),
\label{eq:teta_l}
\end{eqnarray}
where $\vec{l}/\eta$ and $k_z$ are the components perpendicular and parallel to the $\hat{z}$ axis of $\textbf{k}$.

\subsection{The $ <\theta (\vec{x_1})^3 \theta (\vec{x_2})>_c $ statistic for the OV effect}
\label{sec:TrikOV}
For the calculation of the $ <\theta (\vec{x_1})^3 \theta (\vec{x_2})>_c $ statistic for the Ostriker-Vishniac effect, $ <\theta_{ov} (\vec{x_1})^3 \theta_{ov} (\vec{x_2})>_c $, we must compute the trispectrum $T_{ov}(\vec{l}_1, \vec{l}_2, \vec{l}_3, \vec{l}_4)$, which is defined by 
\begin{eqnarray}
<\tilde{\theta}_{ov} (\vec{l}_1) \tilde{\theta}_{ov} (\vec{l}_2) \tilde{\theta}_{ov} (\vec{l}_3) \tilde{\theta}_{ov} (\vec{l}_4)>_c & = & (2\pi)^2 T_{ov}(\vec{l}_1, \vec{l}_2, \vec{l}_3, \vec{l}_4) \delta(\vec{l}_1 + \vec{l}_2 + \vec{l}_3 + \vec{l}_4),
\label{eq:deftri}
\end{eqnarray}
where $\widetilde{\theta}_{ov} (\vec{l})$ is given by equation (\ref{eq:teta_l}). In order to do so, we will use the Limber's approximation \citep{Kai92, Cas04}. It consists in considering $\theta_{ov} (\vec{x})$ as the sum of the contributions $\Delta\theta_{ov} (\vec{x})$ from many shells centered at $\eta_0$ and of width $\Delta\eta$ which is much larger than the relevant wavelengths. These conditions can be expressed as $\Delta\eta / \eta_{rc} \ll 1$ and $\Delta\eta / \eta \gg \varphi$ where $\varphi$ represents the angular scale of interest. Then, according to equation (\ref{eq:teta_l}), the Fourier transform of  $\Delta\theta_{ov} (\vec{x})$ is
\begin{eqnarray}
\Delta\theta_{ov} (\vec{l}) = -\int_{\eta_0 - \frac{\Delta\eta}{2}}^{\eta_0 + \frac{\Delta\eta}{2}}d\eta
\frac{g(\eta)}{c\eta^2} \int \frac{dk_z}{2 \pi}e^{-i k_z \eta} \widetilde{q}_z \Big(\frac{\vec{l}}{\eta}, k_z, \eta \Big).
\label{eq:limber}
\end{eqnarray}
\indent If $g(\eta)$ and $\widetilde{q}_z(\frac{\vec{l}}{\eta}, k_z, \eta)$ vary slowly with the comoving distance $\eta$, they can be considered constant within each shell. Then, one can obtain that
\begin{eqnarray}
\Delta\theta_{ov} (\vec{l}) = -\frac{g(\eta_0)\Delta\eta}{c\eta_0^2} \int \frac{dk_z}{2 \pi}e^{-i k_z \eta_0} \widetilde{q}_z \Big(\frac{\vec{l}}{\eta_0}, k_z, \eta_0 \Big)j_0 \Big(k_z \frac{\Delta\eta}{2} \Big),
\label{eq:delta_teta_l}
\end{eqnarray}
where $j_0(x) = \sin(x)/x$.\newline \indent
In the Limber's approximation, the fields defined within different shells are, by assumption, statistically independent. Therefore, the trispectrum of the OV effect is the sum of the independent trispectra generated in all the shells.\newline \indent
For the calculation of these trispectra, we need to compute $<\widetilde{q}_z(\frac{\vec{l}_1}{\eta}, k_{z1}, \eta)...\widetilde{q}_z(\frac{\vec{l}_4}{\eta}, k_{z4}, \eta)>$. In order to do so, we use the fact that, in the linear regime, $\delta (\textbf{r})$ can be considered as a Gaussian random field, and use the Wick's theorem. \newline \indent
The presence of $j_0(k_z \frac{\Delta\eta}{2})$ in equation (\ref{eq:delta_teta_l}) implies that the contribution to the effect comes mainly from $k_z < 1/\Delta \eta$. In addition, the condition  $\Delta\eta / \eta \gg \varphi$ means that $\vec{l}/\eta \gg 1/\Delta \eta$. Therefore, as stated above, the main contribution to the OV effect comes from Fourier modes of $\widetilde{q}_z(\textbf{k}, \eta)$ nearly perpendicular to the line of sight. Then, we can approximate $\widetilde{q}_z(\frac{\vec{l}}{\eta}, k_z, \eta) \approx \widetilde{q}_z(\frac{\vec{l}}{\eta}, 0, \eta)$ and perform the integration over $k_z$ in equation (\ref{eq:delta_teta_l}) in a very easy way.\newline \indent
Finally, approximating the sum of the trispectra from all the shells to an integral, one can obtain that
\begin{eqnarray}
T_{ov}(\vec{l_1} , \vec{l_2} , \vec{l_3} , \vec{l_4}) =
\frac{1}{8}\int_0^{\eta_0}d\eta
\frac{g(\eta)^4}{\eta^6}\Big(\frac{a\dot{D}(\eta)D(\eta)}{D_0^2}\Big)^4f\Big(\vec{\frac{l_1}{\eta}},
\vec{\frac{l_2}{\eta}}, \vec{\frac{l_3}{\eta}},
\vec{\frac{l_4}{\eta}}\Big),
\label{eq:Tov}
\end{eqnarray}
where
\begin{eqnarray}
f(\vec{k}_1, \vec{k}_2, \vec{k}_3, \vec{k}_4) & = & \int \frac{d\textbf{k}}{(2\pi)^3} \sum_{i,j,k
= 1}^4 P(a)P(b)P(c)P(d)k_z^4 \nonumber \\
& & \times \Big(\frac{1}{a^2}-\frac{1}{b^2}\Big)\Big(\frac{1}{a^2}-\frac{1}{c^2}\Big)
\Big(\frac{1}{d^2}-\frac{1}{b^2}\Big)\Big(\frac{1}{d^2}-\frac{1}{c^2}\Big) (1-\delta_{ij})(1-\delta_{ik})(1-\delta_{jk}),
\nonumber \\ & &
\end{eqnarray}
$a=|\textbf{k}|$, $b=|\textbf{k} - \vec{k}_k|$, $c=|\textbf{k} + \vec{k}_i|$ and $d=|\textbf{k} + \vec{k}_i + \vec{k}_j|$. Notice that equation (\ref{eq:Tov}) differs from equation (60) of \cite{Cas04} by the combinatorial factor $1/4!$.\newline \indent
Now, if we define
\begin{eqnarray}
<\theta_{ov} (\vec{x}_1)^3 \theta_{ov} (\vec{x}_2)>_c \equiv \int\frac{d\vec{l}^2}{(2\pi)^2} e^{i\vec{l}\cdot (\vec{x}_2 - \vec{x}_1)} C_{ov}^{\theta^3 \theta}(l),
\label{eq:definition}
\end{eqnarray}
we have
\begin{eqnarray}
C_{ov}^{\theta^3 \theta}(l) & = & \int\frac{d\vec{l}^2_1}{(2\pi)^2} \int\frac{d\vec{l}^2_2}{(2\pi)^2}T_{ov}(\vec{l}_1, \vec{l}_2 - \vec{l}_1, -\vec{l}_2 -\vec{l} , \vec{l}).
\label{eq:C3}
\end{eqnarray}
So, considering equation (\ref{eq:Tov}), we obtain
\begin{eqnarray}
C_{ov}^{\theta^3 \theta}(l) & = & \frac{3}{2}\int_0^{\eta_{rc}}d\eta
\frac{g(\eta)^4}{c^4\eta^2} \Big(\frac{a\dot{D}(\eta)D(\eta)}{D_0^2}\Big)^4 \int \frac{d\vec{k}_1}{(2\pi)^2} \frac{d\vec{k}_2}{(2\pi)^2} \frac{d\textbf{k}}{(2\pi)^3}k_z^4 \nonumber \\
& & \times P(a')P(b')P(c')P(d') \Big(\frac{1}{a'^2} - \frac{1}{b'^2}\Big)^2 \Big(\frac{1}{c'^2} - \frac{1}{d'^2}\Big)^2,
\label{eq:C32}
\end{eqnarray}
where $a' = \vert \textbf{k} \vert$, $b' = \vert \textbf{k} - \vec{l}/\eta \vert$, $c' = \vert \textbf{k} - \vec{k}_1 \vert$, and $d' = \vert \textbf{k} - \vec{k}_2 \vert$.

\subsection{Discussion}
\label{sec:Discuss}

The amplitude and shape of $C^{\theta^3\theta}_{ov} (l)$ depends on the reionization history of the universe (contained in the visibility function $g(\eta)$). There is not yet certainty about the actual shape of $g(\eta)$; in fact, the aim of this work is to contribute to answering this question. However, in order to show the kind of information that can be obtained from the $<\theta (\hat{n})^3 \theta (\hat{m})>_c$ statistic, in this section we calculate $C^{\theta^3\theta}_{ov} (l)$ making use of a very simple model for $g(\eta)$.\newline
\indent In order to do so, we use the flat-$\Lambda CDM$ cosmological model with the cosmological parameters given by ($\Omega_m, \Omega_b, h, n_s, \tau, \sigma_8$) = (0.24, 0.04, 0.73, 0.95, 0.09, 0.74), in agreement with the three years WMAP results \citep{Spe06}. For the transfer function we make use of the fitting form given by \cite{BBKS86}. We assume a mass fraction of hydrogen and helium of $75 \%$ and $25 \%$, respectively, with simultaneous hydrogen and helium reionizations.\newline 
\begin{figure}
\centering\includegraphics[width=16cm]{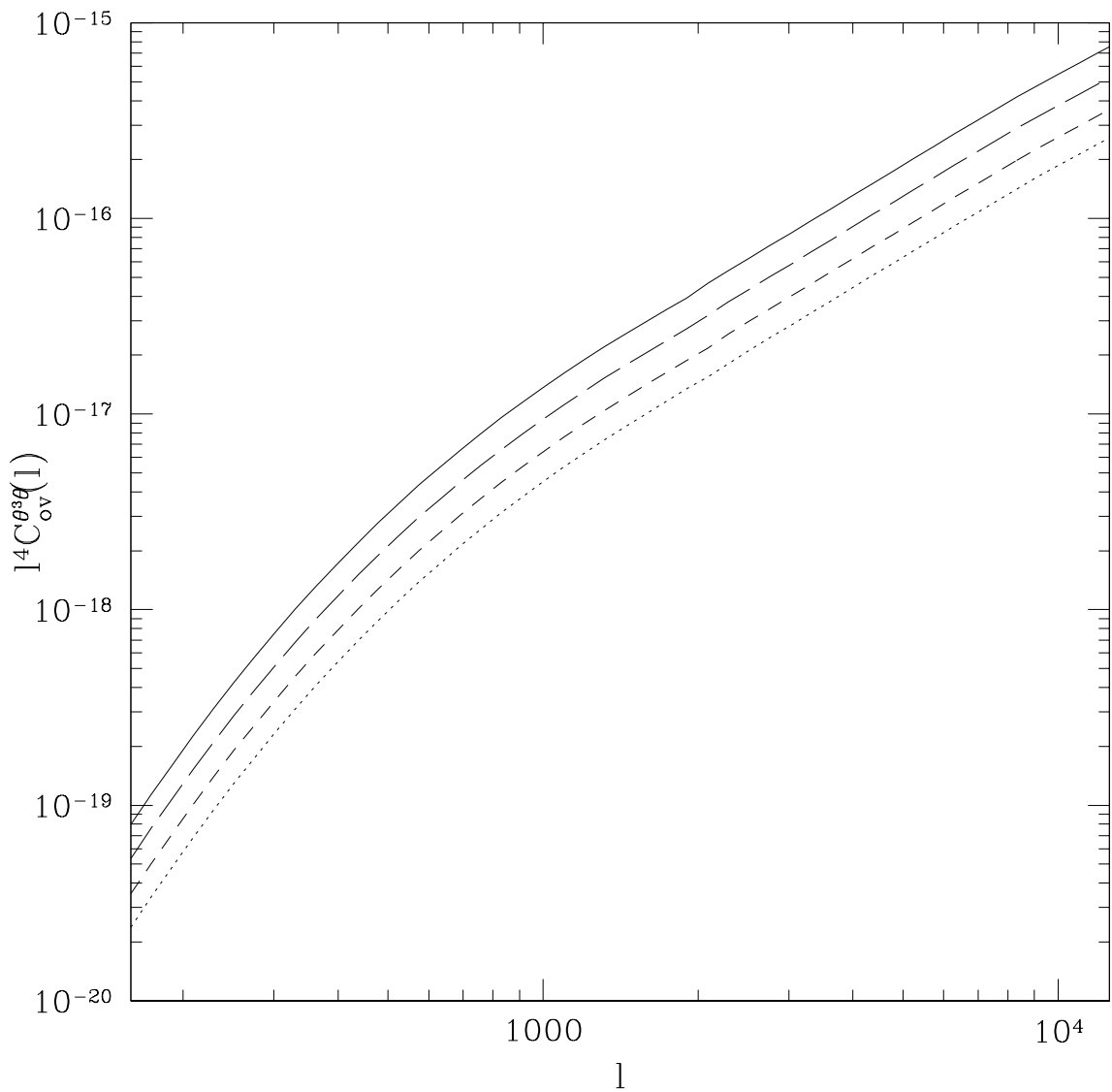}
\caption{$l^4C^{\theta^3\theta}_{ov}(l)$ for four two-stage reionization scenarios. The solid, long-dashed, dashed, and dotted lines, represent reionization histories with $(x^0_e, z_{ri}) = (0.8, 13.3)$, $(0.6, 15.1)$, $(0.4, 18.4)$, and $(0.2, 27)$, respectively. In the four cases we have assumed $\tau = 0.09$ and $\sigma_8 = 0.74$.} \label{fg:figura}
\end{figure}
\indent For the reionization history, we assume a very simple two-stage model. In this model the universe is partially reionized at redshift $z_{ri}$, with an ionization fraction of $x^0_e$. After that, at $z \leq 7$, the universe is completely ionized, i.e.,
\begin{eqnarray}
x_e(z) = \left\{
                           \begin{array}{cc}
         0 & z > z_{ri}  \\
         x^0_e & z_{ri} > z > 7  \\
         1 & z \leq 7.
                           \end{array}
                   \right.
\end{eqnarray}
\indent In figure (\ref{fg:figura}), we show the calculation of $C^{\theta^3\theta}_{ov} (l)$ for four different reionization scenarios. The solid, long-dashed, dashed, and dotted lines, represent reionization histories with $(x^0_e, z_{ri}) = (0.8, 13.3)$, $(0.6, 15.1)$, $(0.4, 18.4)$, and $(0.2, 27)$, respectively. Notice that all these scenarios have the same optical depth to Thompson scattering at the last scattering surface, $\tau = 0.09$ and the same $\sigma_8 = 0.74$.\newline \indent
We can see that, at angular scales relevant for the upcoming high resolution CMB experiments ($10^3 \lesssim l \lesssim 10^4$), $C^{\theta^3\theta}_{ov} (l)$ varies by approximately a factor of $3$ between the comparatively early ($z_{ri} = 27$) and late ($z_{ri} = 13.3$) reionization scenarios. This result suggests that, eventhough $<\theta (\hat{n})^3 \theta (\hat{m})>_c$ also depends on other cosmological parameters like $\tau$ and $\sigma_8$, it would be sensitive enough to provide us with new information about the history of reionization of the universe when complemented with other independent measurements. However, before arriving to any conclusion about the relationship between $C^{\theta^3\theta}_{ov} (l)$ and the physical parameters involved in reionization, our calculation must include more realistic models for the reionization process, and also consider the non-linear behaviour of the kSZ effect.

\section{Conclusions}
\label{Conclusiones}

We propose non-Gaussianity to separate the kSZ effect from the lensing effect in CMB maps. Specifically, we show that a particular shape of the trispectrum, $<\theta^3(\hat{n})\theta(\hat{m})>_c$, cancels for weakly lensed primary CMB anisotropies providing a very good tool for measuring the kSZ effect.\newline
\indent Measurements of the amplitude of the kSZ signal complement measurements of the optical depth and provide new insights into the reionization history of the universe. We show that reionization histories that produce similar low $l$ $<EE>$ signal have very different kSZ signatures. In addition to yield a new insight into the history of reionization and structure formation of the universe, $<\theta^3(\hat{n})\theta(\hat{m})>_c$ could become very useful for quantifying and controlling the kSZ contamination of lensing statistics.\newline
\indent While our calculation is based on a linear theory estimate and needs further investigation with full numerical simulations of reionization, we suspect that they will confirm our basic conclusion that kSZ measurements will shed more light on the dark ages.\newline

\acknowledgments

M.A.R. acknowledges financial support from a CONICYT Fellowship and from Princeton University. M.A.R. also thanks professors Andreas Reisenegger, Leopoldo Infante and Hern\'{a}n Quintana for useful conversations. D.N.S. acknowledges support from NSF PIRE program and NASA ATP program. D.N.S. thanks CTIO for its hospitality during Fall 2004 when this work was initiated. Finally, M.A.R. and D.N.S. appreciate the comments of the referee, Olivier Dor\'e, which allowed them to substantially improve the final version of this work.

\end{document}